\documentclass[12pt]{article}
\usepackage{latexsym,epsfig}
\usepackage{amsmath}
\usepackage{amsfonts}
\usepackage{graphicx}
\newtheorem{lem} {Lemma}

\newtheorem{cor} [lem]{Corollary}
\newtheorem{theo}[lem]{Theorem}

\newcommand{\qed}{\hfill\rule{1ex}{1ex}}

\newcommand{\eps}{\varepsilon}

\newenvironment{pf}{\paragraph{Proof:}}{\qed\\\medskip}

\newcommand{\fname}[1]{\textnormal{\textsf{#1}}}

\newcommand{\join} {\fname{join}}
\newcommand{\Evert}{\fname{evert}}
\newcommand{\Link}{\fname{link}}
\newcommand{\Cut}{\fname{cut}}
\newcommand{\Attach}{\fname{attach}}
\newcommand{\Detach}{\fname{detach}}
\newcommand{\Expose}{\fname{expose}}
\newcommand{\Create}{\fname{create}}
\renewcommand{\Join}{\fname{join}}
\newcommand{\Select}{\fname{select}}
\newcommand{\Split}{\fname{split}}
\newcommand{\Destroy}{\fname{destroy}}
\newcommand{\weight}{\ensuremath{\mathit{weight}}}

\newcommand{\Mnode}{\ensuremath{m}}

\newcommand{\toptree}[1]{\ensuremath{\mathcal{#1}}}
\newcommand{\Top}{\toptree{R}}
\newcommand{\boundarysym}{\ensuremath{\partial}}

\newcommand{\abs}[1]{\ensuremath{\left\lvert{#1}\right\rvert}}
\newcommand{\Set}[1]{\ensuremath{\left\{{#1}\right\}}} 

\newcommand{\boundary}[1]{\ensuremath{\boundarysym{#1}}}

{\end{list}}

\newenvironment{procedures}
{%
\begin{list}{}
{%
\setlength\itemsep{0.05in}%
\setlength\parsep{0in}%
\setlength\itemindent{\labelsep}%
\setlength\labelwidth{\parindent}%
\setlength\leftmargin{\parindent}%
\setlength\rightmargin{0in}%
\setlength\topsep{0.05in}%
}%
\raggedright%
}%
{\end{list}
}

\begin{document}

\title{%
  Maintaining Information in Fully-Dynamic Trees with Top Trees%
  \footnote{%
    This paper includes work presented at ICALP'97~\cite{AHLT97} and
    SWAT'00~\cite{AHT00}.
    }%
  }%

\author{%
  Stephen Alstrup%
  \thanks{%
    The IT University of Copenhagen, Glentevej 67, DK-2400, Denmark. %
    \mbox{E-Mail:}~\texttt{stephen@it-c.dk} %
    } %
  \and %
  Jacob Holm%
  \thanks{%
    The IT University of Copenhagen, Glentevej 67, DK-2400, Denmark. %
    \mbox{E-Mail:}~\texttt{jholm@it-c.dk} %
    } %
  \and %
  Kristian de Lichtenberg%
  \thanks{%
    Fogedmarken 12, 2tv, DK-2200 Copenhagen, Denmark. %
    \mbox{E-Mail:}~\texttt{kdl@k2.dk} %
    Work done while master's student at University of Copenhagen.%
    } %
  \and %
  Mikkel Thorup%
  \thanks{%
    AT\&T Labs--Research, %
    \mbox{E-Mail:}~\texttt{mthorup@research.att.com} %
    } %
  }%

\date{\vspace{-1cm}}
\bibliographystyle{abbrv}

\maketitle

\begin{abstract}  
  We introduce top trees as a design of a new simpler interface for data
  structures maintaining information in a fully-dynamic forest. We
  demonstrate how easy and versatile they are to use on a host of
  different applications. For example, we show how to maintain the
  diameter, center, and median of each tree in the forest. The forest
  can be updated by insertion and deletion of edges and by changes to
  vertex and edge weights. Each update is supported in $O(\log n)$
  time, where $n$ is the size of the tree(s) involved in the
  update. Also, we show how to support nearest common ancestor queries
  and level ancestor queries with respect to arbitrary roots in
  $O(\log n)$ time.  Finally, with marked and unmarked vertices, we
  show how to compute distances to a nearest marked vertex. The later
  has applications to approximate nearest marked vertex in general
  graphs, and thereby to static optimization problems over shortest
  path metrics.
  
  Technically speaking, top trees are easily implemented either
  with Frederickson's topology trees [Ambivalent Data Structures for
  Dynamic {2-Edge-Connectivity} and $k$ Smallest Spanning Trees,
  \emph{SIAM J. Comput.} 26 (2) pp.\ 484--538, 1997] or with Sleator and
  Tarjan's dynamic trees [A Data Structure for Dynamic Trees. \emph{J.
    Comput. Syst. Sc.}  26 (3) pp.\ 362--391, 1983].  However, we
  claim that the interface is simpler for many applications, and
  indeed our new bounds are quadratic improvements over previous
  bounds where they exist.
\end{abstract}

\section{Introduction} \label{sec:int}
In this paper, we introduce top trees as a new simpler interface for data
structures maintaining information in a fully-dynamic forest. Here
fully-dynamic means that edges may be both inserted and deleted. The
information could be, say, the diameter of each tree in the forest. However,
if the tree is a minimum spanning tree of a dynamic graph, the information
could help changing the minimum spanning tree as the graph changes.

Technically speaking, top trees are easily implemented either with 
Frederickson's topology trees \cite{Frederickson97} or with Sleator and
Tarjan's dynamic trees \cite{ST83}. The contribution of top trees is the
{\em design\/} of an interface providing users with easier access to the full
power of these advanced techniques.

Targeting a broad audience of potential users, the bulk of this paper
is like a tutorial where we demonstrate the flexibility of top trees in
different types of applications:
\begin{itemize}
\item We re-derive some of the classic applications from 
\cite{Frederickson97,ST83}, e.g., finding the maximum weight of
a given path.
\item We improve some previous bounds. More specifically, we show
how to maintain the centers and medians of trees in a dynamic
forest in $O(\log n)$ time per updates. The previous bounds were 
$O(\log^2 n)$ time \cite{CN96,APP96}. 
\item We consider problems that appear not to have been studied before
for a dynamic forest. For example, we show how to maintain the
diameters of trees in a dynamic forest. We also show how to answer
level ancestor and nearest common ancestor queries with respect to
arbitrary roots.  Finally, with marked and unmarked vertices, we show
how to compute distances to a nearest marked vertex. In all of these
cases, we support both updates and queries in logarithmic time. The
marking result has applications to approximate nearest marked vertex
in general graphs, and thereby to static optimization problems over
shortest path metrics.
\end{itemize}  
We note that finding medians and centers is more difficult than, e.g.,
finding the minimum edge on a given path because they are
``non-local'' properties. Here, by a \emph{local property} we mean
that if an edge or a vertex has the property in a tree, then it has
the property in all subtrees it appears in.  Local properties lend
themselves nicely to bottom-up computations, whereas non-local
properties tend to be more challenging. Building on top of our top
trees, we present here a quite general technique for dealing with
non-local properties.

We implement our top trees with Frederickson's topology
trees~\cite{Frederickson97}, which we in turn implement with Sleator
and Tarjan's \emph{st-trees}~\cite{ST83}. The implementation of
topology trees with st-trees was not known. It has the interesting
consequence that the simple amortized version of st-trees gives a
simple amortized version of topology trees.

We note that since top trees were originally announced \cite{AHLT97},
they have found applications in other
works~\cite{GKT01,HLT01,Tho01}. All these applications rely on results
presented in this paper. Also, our specific result for dynamic tree
diameters has found its own application in \cite{NGP01}.

\subsection{Preliminaries}
Most of this paper concerns a forest of trees, which means that
if vertices $v$ and $w$ are connected, they are connected by
a unique path, which we shall denote $v\cdots w$.

When we talk about an edge $(v,w)$,
on an implementation level, we often really think of an identifier $e$ of the 
undirected edge with end-points $v$ and $w$. Via arrays, the end-points
can be found from the identifier $e$ in constant time. However, other
information can also be associated with $e$ such as its successor and 
predecessor in the incidence lists around $v$ and $w$.

\subsection{Contents}
The paper is organized as follows. In \S~\ref{sec:toptrees} we
introduce top trees and solve the diameter problem.  In
\S~\ref{secfind} we present our technique for non-local problems,
and solve the center and median problems.  In \S~\ref{lem:method}
we discuss the advantages and limitations of using top trees relative
to other data structures for dynamic trees.  In \S~\ref{gen} we
mention some generalizations of top trees used in later papers.
Finally, in \S~\ref{sec:implement} we implement top trees with
topology trees and topology trees with st-trees. Finally, we have some
concluding remarks in \S~\ref{sec:conclussion}.

\section{Top Trees} \label{sec:toptrees}
A top tree is defined based on a pair consisting of a tree $T$ 
and a set $\boundary{T}$ of at most 2 vertices from $T$, called \emph{external
  boundary vertices}.  Given $(T,\boundary{T})$, any subtree $C$ of
$T$ has a set $\boundarysym_{(T,\boundary{T})} C$ of \emph{boundary
  vertices} which are the vertices of $C$ that are either in
$\boundary{T}$ or incident to an edge in $T$ leaving $C$. Here, by a
\emph{subtree} of an undirected tree, we mean any connected subgraph.
The subtree $C$ is called a \emph{cluster} of $(T,\boundary{T})$ if it
has at least one edge and
at most two boundary vertices. Then $T$ is itself a cluster with
$\boundarysym_{(T,\boundary{T})}T=\boundary{T}$. 
Also, if $A$ is a subtree of $C$,
$\boundarysym_{(C,\boundarysym_{(T,\boundary{T})}C)}
A=\boundarysym_{(T,\boundary{T})} A$, so $A$ is a cluster of
$(C,\boundarysym_{(T,\boundary{T})}C)$ if and only if $A$ is a cluster
of $(T,\boundary{T})$. Since $\boundarysym_{(T,\boundary{T})}$ is a
canonical generalization of $\boundarysym$ from $T$ to all subtrees of
$T$, we will use $\boundarysym$ as a shorthand for
$\boundarysym_{(T,\boundary{T})}$ in the rest of the paper.

\begin{figure}
  \begin{center}
    \leavevmode\epsfig{file=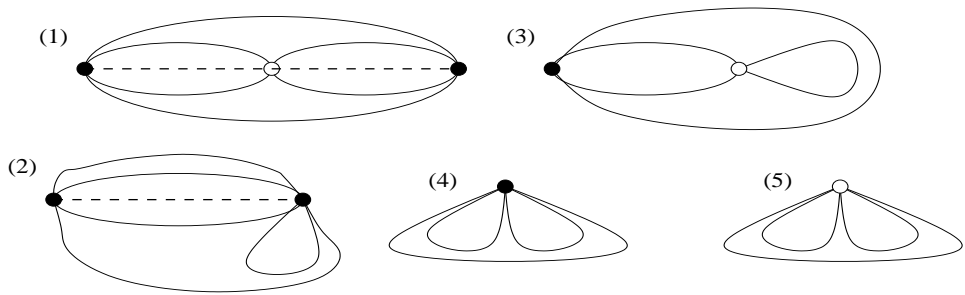,width=5in,height=1.5in}
  \end{center}
  \caption{The cases of joining two neighboring clusters into one. The
    $\bullet$ are the boundary vertices of the joined cluster and the
    $\circ$ are the boundary vertices of the children clusters that did
    not become boundary vertices of the joined cluster. Finally the
    dashed line is the cluster path of the joined cluster.}
  \label{fig:top}
\end{figure}

A \emph{top tree} \toptree{R} over $(T,\boundary{T})$ is a binary tree
such that:
\begin{enumerate}
\item The nodes of \toptree{R} are clusters of $(T,\boundary{T})$.
\item The leaves of \toptree{R} are the edges of $T$.
\item Sibling clusters are {\em neighbors\/} in the sense that
  they intersect in a single vertex, and then their parent cluster is their
  union (see Fig.~\ref{fig:top}).
\item The root of \toptree{R} is $T$ itself.
\end{enumerate}
A tree with a single vertex has an empty top tree.  
The basic philosophy is that clusters are induced by their edges, 
the vertices only being included as their end-points. This is why clusters 
need at least one edge, and we note that neighboring
clusters are induced by disjoint edge sets inducing a common vertex.

We will sometimes refer to the tree $T$ as the {\em underlying\/} tree
to differentiate it from the {\em top\/} tree $\toptree{R}$.

The top trees over the trees in our underlying forest are maintained under
the following {\em forest updates}:
\begin{procedures}
\item[$\Link((v,w))$] where $v$ and $w$ are in different trees, links
  these trees by adding the edge $(v,w)$ to our dynamic forest.
\item[$\Cut(e)$] removes the edge $e$ from our dynamic forest.  
\item[$\Expose(v,w)$] where $v$ and $w$ are in the same tree $T$,
  makes $v$ and $w$ the external boundary vertices of $T$. Moreover,
  $\Expose$ returns the new root cluster of the top tree over $T$. 

  $\Expose$ can also be called with zero or one vertices as argument
  if we want less than two external boundary vertices. If $\Expose$ is
  called with zero arguments, as $\Expose()$, it does not return a root
  cluster. This is because there may be multiple trees, and without an
  argument, $\Expose$ cannot know what tree we are interested in.
  Finally, it is guaranteed that $\Expose()$ does not change the
  structure of the top trees. It only affects some of the boundaries
  of the clusters in the top trees.
\end{procedures}
In general, $\Link$ and $\Cut$ make the set of external boundary
vertices for the resulting trees empty. 
To accommodate these forest updates, the top trees are changed
by a sequence of local {\em top tree modifications\/} described below.
During these modifications, we will temporarily accept a {\em partial\/} top
tree whose root cluster may not be a whole underlying tree $T$ but
just a cluster of $T$.
\begin{procedures}
\item[$e:=\Create()$] creates a top tree with a single cluster $e$
  which is just an edge.
\item[$C:=\Join(A,B)$] where $A$ and $B$ are neighboring 
  root clusters of two top trees $\toptree{R}_A$ and $\toptree{R}_B$.
  Creates a new cluster $C=A\cup B$ and makes it the common root of
  $A$ and $B$, thus turning $\toptree{R}_A$ and $\toptree{R}_B$ into a
  single new top tree $\Top_C$.  Finally, the new root cluster $C$ is
  returned.
\item[$\Split(C)$] where $C$ is the root cluster of a top tree
$\Top_C$ and has children $A$ and $B$. Deletes $C$, thus turning
$\Top_C$ into the two top trees $\Top_A$ and $\Top_B$. Finally, the
root clusters of $\Top_A$ and $\Top_B$ are returned.
\item[$\Destroy(e)$] eliminates the top tree consisting of edge $e$.
\end{procedures}

\subsection{Discipline for modifying top trees}\label{top-dicipline}
Top tree modifications have to be applied in the following
order:
\begin{enumerate}
\item\label{st:split} First, top-down, we perform
a sequence of $\Split$s. 
\item\label{st:destroy} Then we $\Destroy$ the clusters of some edges.
\item\label{st:update} Then we update the forest.
\item\label{st:create} Then we $\Create$s clusters of some edges.
\item\label{st:join} Finally, with $\Join$s, we recreate the top tree 
bottom-up.
\end{enumerate}
The above order implies that when we do a $\Split$ or $\Join$, we
know that all parts of the underlying forest is partitioned into base
clusters.

It is an important rule that {\em a forest update may not change any
current cluster}. Here, a cluster is changed by a forest update if the
update changes its set of edges or its set of boundary vertices. To
appreciate the latter, consider an update $\Expose(v)$. This update
only changes clusters with $v$ an interior vertex. A cluster in which
$v$ is already a boundary vertex is not changed. Satisfying the rule
means that when we get to the update in step \ref{st:update}, the
previous steps \ref{st:split}--\ref{st:destroy} should have eliminated
all clusters that would be changed by the update.

It is often natural to perform a {\em composite sequence of updates\/} in step
\ref{st:update}. For example, if dealing with a spanning tree $T$, we
might want to swap one tree edge $(v_1,w_1)$ with another edge
$(v_2,w_2)$. If we do 
$\left\langle\Cut((v_1,w_1));\Link((v_2,w_2)\right\rangle$
as a composite update rather than as two separate updates, we avoid
dealing with a temporary forest when we do the top tree modifications in
steps \ref{st:split}--\ref{st:destroy} and
\ref{st:create}--\ref{st:join}.

In this paper, we are going to show the following result:
\begin{theo}\label{thm:top-maintain}\sloppy
  For a dynamic forest we can maintain top trees of height $O(\log n)$
  supporting each $\Link$, $\Cut$, or $\Expose$ with a sequence of
  $O(1)$ $\Create$ and $\Destroy$, and $O(\log n)$ $\Join$ and
  $\Split$.  These top tree modifications are identified in $O(\log n)$ time.  The
  space usage of the top trees is linear in the size of the dynamic
  forest. For a composite sequence of $k$ updates, each of the
  above bounds are multiplied by $k$.
\end{theo}
The proof of Theorem~\ref{thm:top-maintain} is deferred to
\S~\ref{sec:implement}. Until then, the focus will be on
applications of top trees. 

\subsection{Top trees generalize balanced binary search trees}\label{binary}
Put in perspective, our top trees are natural generalizations of
standard balanced binary trees over dynamic collections of lists that
may be concatenated and split.  In the balanced binary trees, each
node represents a segment of a list, which in top terminology is just a
special case of a cluster.  Standard implementations for balanced
binary trees also ascertain that the height is $O(\log n)$, and that
each concatenation and  split can be done by $O(\log n)$ local modifications.


\subsection{Top tree terminology} 
If a vertex in a cluster is not a boundary vertex, it is
\emph{internal} to that cluster.  If a
cluster $C$ has two boundary vertices $a$ and $b$, we call $C$ a
\emph{path cluster} and $a\cdots b$ the \emph{cluster path} of $C$,
denoted $\pi(C)$.  If $C$ has only one boundary vertex $a$, $C$ is
called a \emph{point cluster} and then $\pi(C)=a$.  Note that if $A$
is a child cluster of $C$ and $A$ shares an edge with $\pi(C)$, then
$\pi(A)\subseteq\pi(C)$, and then we call $A$ a \emph{path child} of
$C$. In terms of boundary vertices, if $C$ has children $A$ and $B$,
$A$ is a path child of $C$ if and only if $\abs{\boundary{C}}=2$ and
either $\boundary{A}=\boundary{C}$ (Fig.~\ref{fig:top}\,(2)) or
$\boundary{C}\subset\boundary{A}\cup\boundary{B}$
(Fig.~\ref{fig:top}\,(1)). 

\subsection{Representation and usage of top trees} A top tree
is represented as a standard binary rooted tree with parent and
children pointers.  The nodes used to represent the top tree are
denoted \emph{top nodes}. The top nodes of the binary tree
represent the clusters, and with each top node is associated the set
of at most two boundary vertices of the represented cluster. With a top leaf we
store the corresponding edge.  With an internal top node is stored how it is
decomposed into its children (c.f.~Fig.~\ref{fig:top}). Thus,
considering the information descending from a top node, we can
construct the cluster it represents.
Finally, from each vertex $v$, there is a
pointer to the smallest cluster $C(v)$ that $v$ is internal to, or to
the root cluster containing $v$ if $v$ is an external boundary
vertex. 

Following parent pointers from $C(v)$, we can find the root, 
$top\_root(v)$, of the top tree over the underlying tree $T$ containing $v$. 
In the case of a forest, two vertices $v$ and $w$ are in the same 
underlying tree if and only if $top\_root(v)=top\_root(w)$.
With top trees of logarithmic height as in Theorem \ref{thm:top-maintain},
we identify $top\_root(v)$ in $O(\log n)$ time.


An \emph{application} of the top tree data structure, such as
maintaining diameters, centers, or medians, has direct access to the
above representation, and will typically associate some extra
information with the top nodes. The application employs an 
\emph{implementation} of top trees, which is an algorithm like the one
described in Theorem~\ref{thm:top-maintain}, converting each $\Link$,
$\Cut$, or $\Expose$ into a sequence of $\Split$s and $\Join$s on the
top trees.  In connection with each $\Join$ and $\Split$ the
application is notified and given pointers to the top nodes
representing the involved clusters. The application can then update
its information associated with these top nodes. We note that a top
tree may only be modified with $\Split$ and $\Join$. This discipline
is important if we have several applications running over the same top
trees, each maintaining its own information as $\Split$s and $\Join$s
are performed. Typically, $\Link$ and $\Cut$ are operations imposed
from the outside whereas $\Expose$ typically is used internally by an
application.

\subsection{Concrete applications}
As a first example, we can now easily derive a main result
from~\cite{ST83}.
\begin{theo}[Sleator and Tarjan]\label{cor:path-query} 
  We can maintain a dynamic collection of weighted trees in $O(\log
  n)$ time per $\Link$ and $\Cut$, supporting queries about the
  maximum weight between any two vertices in $O(\log n)$ time.
\end{theo}
\begin{pf}
  For this application, with each (top node representing a) cluster
  $C$, we store as extra information the maximum weight
  $max\_weight(C)$ on the cluster path $\pi(C)$. For a point-cluster
  $C$, $max\_weight(C)=-\infty$. If a path cluster consists of a
  single edge $e$, $max\_weight(e)$ is just the weight of the
  edge. When a path cluster $C$ is created by a $\Join$,
  $max\_weight(C)$ is the maximum weight stored at its path
  children. When $C$ is $\Split$ or $\Destroy$ed, we just discard the information
  stored with $C$. Now, to find the maximum weight between $v$ and
  $w$, we set $C:=$ $\Expose(v,w)$. Then $\pi(C)=v\cdots w$, and we
  return $max\_weight(C)$.  Since $\Join$ and $\Split$ are supported
  in constant time, the Theorem now follows from
  Theorem~\ref{thm:top-maintain}.
\end{pf}

In the above example, $\Split$ is trivial. To see the relevance of
$\Split$, we consider an extension from~\cite{ST83}.
\begin{theo}[Sleator and Tarjan]\label{cor:path-update} 
  In Theorem~\ref{cor:path-query}, we can also add a common weight $x$
  to all edges on a given path $v\cdots w$ in $O(\log n)$ time.
\end{theo}
\begin{pf}
  For this extension, for each cluster $C$, we introduce a ``lazy''
  weight $extra(C)$ which is to be added to all edges in $\pi(C)$
  in all clusters properly descending from $C$. We note that if
  $C$ is a root cluster, $max\_weight(C)$ is not affected by these
  $extra$-values, so $max\_weight(C)$ is the correct maximal weight
  on $\pi(C)$. In particular, we can still find the maximal weight
  between $v$ and $w$ as $max\_weight(\Expose(v,w))$.

  The addition of $x$ to $v\cdots w$ is now done by calling
  $C:=\Expose(v,w)$ and adding $x$ to $max\_weight(C)$ and to
  $extra(C)$.  Then $\Split(C)$ requires that for each path child $A$
  of $C$, we set $max\_weight(A):=max\_weight(A)+extra(C)$ and
  $extra(A):=extra(A)+extra(C)$.  For $C:=\Join(A,B)$, we set
  $max\_weight(C):=\max\Set{max\_weight(A), max\_weight(B)}$ and
  $extra(C):=0$.  Finally, to find the maximum weight on the path
  $v\cdots w$, we set $C:=\Expose(v,w)$ and return $max\_weight(C)$.
\end{pf}

We will now go beyond \cite{ST83} with a further extension needed
in \cite{Tho01}. 
\begin{theo}\label{cor:tree-weight} 
  In Theorem~\ref{cor:path-update}, we can also ask for the maximum
  weight of the underlying tree containing a vertex $v$ in $O(\log n)$ time.
\end{theo}
\begin{pf}
  Elaborating on the information from the previous two proofs, for
  each cluster $C$, we will maintain a variable $max\_non\_path(C)$
  denoting the maximal weight on an edge in $C$ which is not on the
  cluster path. Assuming this variable, we can find the maximal weight
  of the underlying tree containing $v$, setting $C:=top\_root(v)$ and
  returning $\max\{max\_non\_path(C),max\_weight(C)\}$.

  We maintain the $max\_non\_path$ variables as follows. When the
  cluster of an edge $e$ is $\Create$d, if $e$ has two boundary vertices,
  we set $max\_non\_path(e)=-\infty$. Otherwise, $max\_non\_path(e)$ is
  set to the weight of $e$. When a cluster is joined as
  $C:=\Join(A,B)$, we first set $max\_non\_path(C):=
  \max\{max\_non\_path(A),max\_non\_path(B)\}$. If $C$ is not a path
  cluster but one of its children, say $A$, is a path cluster
  (c.f.~Fig.~\ref{fig:top}(3)),
  then we further have consider weights from the cluster path of $A$,
  setting $max\_non\_path(C):=\max\{max\_non\_path(C),max\_weight(A)\}$.
  We note here that because $A$ was a root cluster, $max\_weight(A)$
  has its correct value, not missing any $extra$-values from 
  at ascending clusters. When clusters are $\Split$ or $\Destroy$ed,
  this has no impact on the $\max\_non\_path$-variables.
\end{pf}

In the rest of this paper, we are more interested in distances than
in maximum weights. Modifying the proof of Theorem \ref{cor:path-query},
for each cluster $C$, we will maintain the length $length(C)$ of the
cluster path.  The length is maintained as the maximum weight except
that if $C$ is created by a $\Join$, $length(C)$ is the sum of
lengths stored with its path children. Thus we have
\begin{lem}\label{lem:dist} 
  In top trees, for each cluster $C$, we can maintain the length,
  denoted $length(C)$, of the cluster path in constant time per local top
  update, hence in $O(\log n)$ time per $\Link$ or $\Cut$. Then the
  distance between two vertices $v$ and $w$ can be found in $O(\log
  n)$ time as $length(\Expose(v,w))$. \qed
\end{lem}

As an interesting new application of top trees, we get the claimed
result for dynamic diameters.
\begin{theo}\label{cor:diameter} 
  We can maintain a dynamic collection of weighted trees in $O(\log
  n)$ time per $\Link$ and $\Cut$, supporting queries about the
  diameter of the tree containing any vertex in $O(\log n)$ time.
\end{theo}
\begin{pf}
  For each cluster $C$, we store its diameter $diam(C)$.
  Moreover, for each of its
  boundary vertices $a\in\boundary{C}$, we store the maximal distance
  $max\_dist(C,a)$ from $a$ to any vertex in $C$. Finally, we maintain
  the cluster length from Lemma \ref{lem:dist}. The variables
  $max\_dist$ and $length$ are auxiliary fields, needed for a fast $\Join$.
  Such carefully chosen extra information is often crucial
  in top tree applications.

  When the cluster of an edge $e$ is $\Create$d, $diam(e)=weight(e)$,
  and for each boundary vertex $v$ of $e$, $max\_dist(e,v)=weight(e)$.
  Now, suppose $C:=\Join(A,B)$, and that $c$ is the common boundary
  vertex of $A$ and $B$. Then we set
  \[diam(C):=\max\Set{diam(A),diam(B),max\_dist(A,c)+max\_dist(B,c)}\]
  Now consider any boundary vertex $a$ of $C$. By symmetry,
  we may assume that if $a$ is not in one of $A$ and $B$, it is not in $B$. 
  Let $c$ be the intersection vertex of $A$ and $B$. Then, if $c \neq a$, 
  \[max\_dist(C,a)=\max\Set{max\_dist(A,a),length(A)+max\_dist(B,c)}\]
  If $c=a$ then  
 \[max\_dist(C,c)=\max\Set{max\_dist(A,c),max\_dist(B,c)}\]
  Thus, $\Create$ and $\Join$ are implemented in constant time. As in 
  the proof of
  Theorem~\ref{cor:path-query}, $\Split$ and $\Destroy$  do 
  not require any action.
  Hence Theorem~\ref{thm:top-maintain} implies that we can maintain
  the above information in $O(\log n)$ time per $\Link$ or $\Cut$. To
  answer a diameter query for a vertex $v$, we set $C:=\Expose(v)$ and
  return $diam(C)$.
\end{pf}

Another illustrative application is the maintenance of nearest marked
neighbors.
\begin{theo}\label{thm:nearest-marked} 
  We can maintain a dynamic collection of trees in $O(\log
  n)$ time per $\Link$ and $\Cut$, or marking and unmarking of a vertex,
  supporting queries about the (distance to) the nearest marked vertex
  of any given vertex in $O(\log n)$ time.
\end{theo}
\begin{pf}
Below, we just focus on finding the distance to the nearest
marked vertex. This is easily extended to also providing the vertex.

For each boundary vertex $a$ of a cluster $C$, we maintain the
distance $mark\_dist(C,a)$ from $a$ to the nearest marked vertex in
$C\setminus\boundary C$.  The reason that why exclude the boundary of $C$ 
from consideration is that a vertex $v$ may appear as boundary
vertex of $\Omega(n)$ clusters, and all these would be affected, if $v$ was
(un)marked. From $mark\_dist(C,a)$ we can easily compute
the distance $mark\_dist^*(C,a)$ from $a$ to the nearest marked vertex in 
$C$ excluding only boundary vertices different from $a$. Then
$mark\_dist^*(C,a)=0$ if $a$ is marked, and 
$mark\_dist^*(C,a)=mark\_dist(C,a)$ if $a$ is unmarked. We also
maintain the cluster path length, $length(C)$, as in Lemma~\ref{lem:dist}.

Given a vertex $u$, to find the distance to the nearest marked vertex,
we simply set $C:=\Expose(u)$, and return
$mark\_dist^*(C,u)$.

To (un)mark a vertex $v$, we first $\Expose$ $v$. As an external
boundary vertex, $v$ has no impact on any $mark\_dist$-value, so we
can freely (un)mark it.

Suppose the cluster $C$ is $\Create$d as an edge $(v,w)$.  Then
$mark\_dist(C,v)$ is the weight of $(v,w)$ if $w$ is marked and not in
the boundary; otherwise, we it to infinity.

Finally, consider $C:=\Join(A,B)$ with $\{c\}=A\cap B$.
Let $a$ be a boundary vertex of $C$. By symmetry,
we can assume that $a$ is in $A$. We now have $mark\_dist(C,a)=$
\[\left\{\begin{array}{ll}
\min\{mark\_dist(A,a),mark\_distB,a)\}&\mbox{if $a=c$}\\
\min\{mark\_dist(A,a),length(A)+mark\_dist(B,c)\}&\mbox{if $a\neq c$ and $c\in\boundary C$}\\
\min\{mark\_dist(A,a),length(A)+mark\_dist^*(B,c)\}&\mbox{if $a\neq c$ and $c\not\in\boundary C$}
\end{array}\right.\]
Thus, we can support both $\Join$ and $\Create$ in constant time, and 
$\Split$ and $\Destroy$  do not require any action. By 
Theorem~\ref{thm:top-maintain}, this completes the proof of Theorem
\ref{thm:nearest-marked}
\end{pf}
\begin{cor} For any positive integer parameter $k$, in a 
fixed undirected graph on $n$ vertices and $m$ edges, in 
$O(kmn^{1/k}\log n)$ expected time we can build an $O(kn^{1+1/k})$ space
data structure, supporting (un)marking of vertices and queries
about stretch $2k-1$ distances to a nearest marked vertex. 
Here stretch $2k-1$ means that the reported distance may be
up to a factor $2k-1$ too long. Both
queries and updates take $O(kn^{1/k}\log n)$ time.
\end{cor}
\begin{pf} In \cite{TZ01}, it is shown how to generate a cover
of edge-induced trees within the above preprocessing bounds so that
each vertex $v$ is in $O(kn^{1/k})$ trees, and if the distance from $v$ to
$w$ is $d$, there is a tree in which the distance is at most
$(2k-1)d$. Now, if a vertex is marked, it is marked in all the trees
containing it, and to find a stretch $2k-1$ distance to a nearest marked 
vertex, we find the shortest distance to a marked vertex over all
the trees.
\end{pf}

The above corollary is interesting because it in \cite{GIV01} is shown
that several combinatorial optimization problems can be approximated
efficiently on metrics with dynamic nearest neighbor. For example,
in the bottle-neck matching problem, where we wish to minimize the furthest
distance between a pair in the matching, we now get a $4k-2$ approximation
in $\tilde O(mn^{1/k})$ expected time. An exact solution currently
requires $\tilde O(mn+n^{2.5})$ time \cite{EK75}.

\section{Non-Local searching}\label{secfind}
We are now going to build a black box on top of our top trees for
maintenance of centers and medians. As discussed in the introduction,
the common feature of centers and medians is that they represent
non-local properties. Here a vertex/edge property is local if it being
satisfied by a vertex/edge in a tree implies that the vertex/edge
satisfies the property in all subtrees containing it.  For example,
being the minimum edge on a given path is a local property. Local
properties lend themselves nicely to bottom-up computations whereas
non-local properties appear to be more challenging.

For our general non-local searching, the application should supply a
function $\Select$ that given the root cluster of a top tree,
selects one of the two children. Recall here that a root cluster
represents the whole underlying tree, which is important when dealing
with non-local properties. Our black box will use $\Select$ to guide a
binary search after a desired edge. More precisely,
the first time $\Select$ is called, it is just given the
root of an original top tree $\Top$. It then $\Select$s one of
the two children. In subsequent iterations, there will be some cluster
$C$ in the original top tree which is the intersection of all
clusters $\Select$ed so far. If $C$ has children $A$ and $B$, the
black box modifies the top tree so that $A$ and $B$ are subsumed
by different children $A^*$ and $B^*$ of the root. Then $\Select$ is
called on the root $C^*=\Join(A^*,B^*)$. If $A^*$ is $\Select$ed, $A$
is the new intersection of all $\Select$ed clusters. Likewise,
if $B^*$ is $\Select$ed, $B$ is the new intersection of all $\Select$ed 
clusters. This way, $\Select$ is used to guide a binary search down
through the original top tree $\Top$. The formal statement of the
result is as follows.
\begin{theo} [Non-Local Search]\label{thm:non-localsearch}
  Starting with the root cluster of a top tree of height $h$ and at most one
  external boundary vertex, after $O(h)$ calls to $\Select$, $\Join$, 
  and $\Split$, there is a unique edge
  $(v,w)$ contained in all clusters chosen by $\Select$, and then
  $(v,w)$ is returned. Subsequently, the top tree is returned
  to its previous state with $O(h)$ calls to $\Join$, and $\Split$.

  If there are two external boundary vertices $x$ and $y$, the above
  selection process will stop with a unique $(v,w)$ edge on the path from
  $x$ to $y$.
\end{theo}
As stipulated in the general interface to top trees, the
implementation behind Theorem~\ref{thm:non-localsearch} will only
manipulate the top tree with $\Join$ and $\Split$ operations.  In our
applications, we will apply Theorem~\ref{thm:non-localsearch} to a top
tree from Theorem~\ref{thm:top-maintain} with height $h=O(\log n)$.
Then the number of calls to $\Join$ and $\Split$ in
Theorem~\ref{thm:non-localsearch} is $O(h)=O(\log n)$.

Theorem~\ref{thm:non-localsearch} will not be proved till
\S~\ref{sec:search-impl}. Before that we demonstrate applications
of Theorem~\ref{thm:non-localsearch} in the dynamic center, median, and
ancestor problems. In these applications, our general approach is to
first decide the information needed for $\Select$, second show how to
make the information available. The external boundary vertices will
only play a role in the ancestor application in \S \ref{sec:ancestor}. 

\subsection{Dynamic center}
For any tree $T$ and
vertex $v$ let $max\_dist(T,v)$ denote the maximal distance from $v$
in $T$. A {\em center\/} is a vertex $v$ minimizing $max\_dist(T,v)$.
\begin{lem}\label{1centerlem} 
  Let $T$ be a tree, and let $A$ and $B$ be neighboring clusters
  with $A\cap B=\Set{c}$ and $A\cup B=T$. If $max\_dist(A,c)\geq 
  max\_dist(B,c)$,   $A$ contains all centers. 
\end{lem}
\begin{pf} \sloppy
  Let $w$ be a vertex in $A$ of maximal distance to $c$. Then
  $dist(c,w)=max\_dist(A,c)=max\_dist(T,c)$. Now, for any $v\in B\setminus A$,
  $max\_dist(T,v)\geq dist(v,w)=dist(v,c)+dist(c,w)=dist(v,c)+max\_dist(T,c)$.
  Since the edge weights are positive, \mbox{$dist(v,c)>0$}, thus
  $max\_dist(T,v)>max\_dist(T,c)$ and $v$ cannot be a center.
\end{pf}

In the dynamic center problem, we maintain a forest under $\Link$ and
$\Cut$ interspersed with queries $center(u)$ requesting the center
of the current tree containing the vertex $u$. We use the top trees from
Theorem \ref{thm:top-maintain}. For each boundary vertex $a$ of
a cluster $C$,  we maintain the maximal distance $max\_dist(C,a)$ 
from $a$ in $C$ as described in the proof of Theorem~\ref{cor:diameter}.
Then $\Link$ and $\Cut$ take $O(\log n)$ time.

To find $center(u)$, we first set $D:=\Expose(u)$ so that $D$ becomes
the current root cluster over the tree containing $u$. The non-local
search of Theorem~\ref{thm:non-localsearch} will start in $D$, but
we need to define $\Select$ given an arbitrary root cluster $C$ with 
children $A$ and $B$, $A\cap B=\Set{c}$. If
$max\_dist(A,c)\geq max\_dist(B,c)$, $\Select$ picks $A$, otherwise it
picks $B$. By Lemma~\ref{1centerlem}, any cluster picked contains all
centers, so, following Theorem~\ref{thm:non-localsearch}, the returned
edge $(v,w)$ contains all centers. Moreover, $\Select$ takes constant
time, so $(v,w)$ is found in $O(\log n)$ time.
To find out if $v$ or $w$ is a center, we compute
$D:=\Expose(v,w)$ in $O(\log n)$ time. Since $D$ coincides with $T$, we
can return $v$ if $max\_dist(D,v)<max\_dist(D,w)$; $w$ otherwise. Hence
we can answer $center(u)$ in $O(\log n)$ time. Thus we conclude
\begin{theo}\label{thm:center}
  The center can be maintained dynamically under \Link{}, \Cut{}
  and $center(u)$ queries in $O(\log n)$ worst case time per operation.
\end{theo}

\subsection{Dynamic median}
Let $T$ be a tree with positive vertex and edge weights.  A
{\em median\/} is a vertex $\Mnode$ minimizing
$\sum_{v\in{}V}(\weight(v)\times dist(v,\Mnode))$ where
$dist(v,\Mnode)$ is the distance from 
$v$ to $\Mnode$ in the tree. For any tree $T$, let $vert\_weight(T)$
denote the sum of the vertex weights of $T$. Our approach to finding
medians is similar to that for centers, but for the median, it is
natural to allow the application to change vertex weights, and this
requires a simple trick.

The simple lemma below is implicit in Goldman~\cite{Goldman71}.
\begin{lem}\label{1medianlem}\sloppy
  Let $(v,w)$ be an edge in the weighted tree $T$, and let $T_v$ and
  $T_w$ be the trees from $T\setminus\Set{(v,w)}$ containing $v$ and
  $w$, respectively.  If~\mbox{$vert\_weight(T_v)=vert\_weight(T_w)$},
  $v$ and $w$ are the only medians in $T$, and if
  $vert\_weight(T_v)>vert\_weight(T_w)$, all medians in $T$ are in
  $T_v$.
\end{lem}

\begin{cor}\label{1mediancor}
  Let $T$ be a tree, and let $A$ and $B$ be neighboring clusters with
  $A\cap B=\Set{c}$ and $A\cup B=T$. Then $vert\_weight(A)\geq 
  vert\_weight(B)$ implies that $A$ contains a median of $T$.
\end{cor}

\begin{pf} 
  Assume that $vert\_weight(A)\geq vert\_weight(B)$. If there exists an
  edge $(c,w)$ in $B$ such that $vert\_weight(T_c)=vert\_weight(T_w)$,
  then by Lemma~\ref{1medianlem}, $c$ and $w$ are (the only) medians
  in $T$ and since $c$ is in $A$ we are done.  Otherwise for any edge
  $(c,w)$ in $B$, $vert\_weight(T_c)\neq vert\_weight(T_w)$. By
  assumption, $vert\_weight(T_c)\geq vert\_weight(A)\geq
  vert\_weight(B)\geq vert\_weight(T_w)$, and thus
  $vert\_weight(T_c)>vert\_weight(T_w)$. Then Lemma~\ref{1medianlem}
  states that all medians of $T$ are in $T_c$, and since this is true
  for any edge $(c,w)$, there must be a median in $A$.
\end{pf}

The above corollary suggests that we should maintain the vertex weight
of each cluster, but this gives rise to a problem; namely that a
single vertex can be contained in arbitrarily many clusters, and a
change in its weight would affect all these clusters. Recall
that we faced a very similar problem for the $mark\_dist$-values in the
proof of Theorem \ref{thm:nearest-marked}, and again we will resort
to ignoring the boundary. 

For each cluster $C$, we only maintain their
``internal weight'' $int\_weight(C)=vert\_weight(C\setminus\boundary{C})$.
We can still derive the real weight $vert\_weight(C)$ as 
$int\_weight(C)+weight(\boundary C)$ in constant time. 

To $\Join$ two clusters $A$ and $B$, $A\cap B=\Set{c}$
into $C$, we add their internal weights plus the weight of $c$ if
$c\not\in\boundary{C}$. To change the weight of a
vertex $v$, we first call $\Expose(v)$. Then $v$ is not internal to
any cluster, and hence no cluster information has to be updated when
we change the weight of $v$.

We can now implement $\Select$ as suggested by Corollary \ref{1mediancor},
choosing the child cluster minimizing $vert\_weight$ in constant
time. Thus we get an edge $(v,w)$ which contains all medians in 
$O(\log n)$ time.

To find a median among $v$ and $w$, we apply
Lemma~\ref{1medianlem}. We $\Cut$ the edge $(v,w)$, and return $v$ if
the (root cluster of the) tree $T_v$ containing $v$ is heavier; otherwise
we return $w$. Before returning $v$ or $w$, we $\Link$ $(v,w)$ back 
in $T$. The $\Link$ and $\Cut$ take $O(\log n)$ time, so we conclude:
\begin{theo}\label{thm:median}
  The median can be maintained dynamically under \Link{}, \Cut{} and
  change of vertex weights in $O(\log n)$ worst case time per
  operation. \qed 
\end{theo}

\subsection{Nearest common ancestors and level ancestors}\label{sec:ancestor}
We will now show how to implement nearest common ancestors and
level ancestors with respect to arbitrary roots. In the
context of unrooted trees, this is done via the two
functions $jump(x,y,d)$, returning the vertex $d$ hops
from $x$ on the path from $x$ to $y$, and $meet(x,y,z)$ returning the
intersection point between the three paths connecting $x$, $y$, and $z$.
With root $r$, the level $\ell$ ancestor of $v$ is 
$jump(r,v,\ell)$, and the nearest common ancestor of $u$ and $v$ is
$meet(u,v,r)$.

To implement $jump$ and $meet$, from Lemma \ref{lem:dist} 
we will use the cluster path length $length(\cdot)$ as well as the general 
distances between vertices. To implement
$jump(x,y,d)$ we first $\Expose$ $x$ and $y$. We now implement
$\Select$ as follows. Let $A$ and $B$ be the children of the root cluster
$C$ with $x\in A$ and $y\in B$. If $length(A)\leq d$, we select $A$; otherwise
we select $B$. At the end, we get an edge, and then we return the
end-point whose distance to $x$ is $d$.

Having implemented $jump$, we compute $meet(x,y,z)$ as 
$jump(z,x,(dist(x,z)+dist(y,z)-dist(x,y))/2)$. Thus we conclude
\begin{theo}\label{cor:jump-meet} 
  We can maintain a dynamic collection of weighted trees in $O(\log
  n)$ time per $\Link$ and $\Cut$, supporting $jump$ and $meet$ 
  queries in $O(\log n)$ time. \qed
\end{theo}

\subsection{Non-Local search implementation}\label{sec:search-impl}
We will now first prove Theorem~\ref{thm:non-localsearch} when there are 
no boundary vertices. First we
will assume that there are no external boundary vertices. Essentially our
search will follow a path down the given top tree $\Top$. As we search
down, we will modify the top tree so as to facilitate calls to
$\Select$, but we will end up restoring it in its original form.  All
modifications for the search are done via $\Split$ and $\Join$, as
stipulated in the general interface to top trees.

Our search consists of $O(\log n)$ iterations $i=0,...$.  At the
beginning of iteration $i$, there will be a ``current''
cluster $C_i$ on depth $i$
in the original top tree $\Top$ which contains exactly the edges that have
been in all clusters selected so far. Thus $C_0$ is the original
root cluster representing an underlying tree $T$. If $C_i$ is a single edge
$(v,w)$, we return $(v,w)$.  Otherwise $C_i$ has children $A_i$ and
$B_i$ in the original top tree.  Then $\Select$ will be presented a
root cluster joining $A^*_i$ and $B^*_i$ such that $A_i\subseteq
A^*_i$, $B_i\subseteq B^*_i$, and $T=A_i^*\cup B_i^*$. That is,
the application-defined $\Select$ will be called as
$\Select(\Join(A^*_i,B^*_i))$. If the
application selects $A^*_i$, we have $C_{i+1}=A_i$ for the next
iteration. Otherwise $C_{i+1}=B_i$.

At the beginning of iteration $i$, we have $C_i$ the root of a top
tree which was the subtree of the original top tree $\Top$ descending
from $C_i$. Besides, for each boundary vertex $a$ of $C_i$, we have
an ``outside'' root cluster $X_a$ with everything from the underlying
tree $T$ that is separated from $C_i$ by $a$. Also, $X_a$ includes
$a$.  Together with $C_i$, the outside root clusters $X_a$ partition the
edges of $T$. For $C_0=T$, we do not have any outside root clusters.

We are done when $C_i$ is a top leaf consisting of a single edge.
Otherwise, we $\Split$ $C_i$ into two children $A_i$ and $B_i$.

To create $A_i^*$, we take all outside root clusters intersecting $A_i$ and
$\Join$ them with $A_i$. If an outside root cluster does not intersect $A_i$,
it intersects $B_i$, and is $\Join$ed with $B_i$ to create $B_i^*$. We then
call the application-defined $\Select$ on $\Join(A_i^*,B_i^*)$.

We now $\Split$ all the newly $\Join$ed clusters so that the root clusters
become $A_i$, $B_i$, and the outside root cluster for each boundary vertex
of $C_i$ from the beginning of
the iteration. By symmetry, we
may assume that $\Select$ picked $A_i$. We then set $C_{i+1}:=A_i$, and
we $\Join$ $B_i$ with all outside root clusters intersecting $B_i$ in
a new maximal outside root cluster.
Finally, we recurse on $C_{i+1}$.

As mentioned, the iterations stop as soon as we arrive at a $C_i$
which is just a single edge $(v,w)$. Since each iteration only
involves a constant number of $\Join$s and $\Split$s, we conclude that
the total number of $\Join$s and $\Split$s is $O(h)$ where $h$ is the
initial height of the top tree. In the end
when we have found $C_i=(v,w)$, we just reverse all $\Join$s and
$\Split$s to restore the top tree in its original form, and return the
edge $(v,w)$.

With a minor modification, the above construction also works in the
presence of a single external boundary vertex. The modification is in
the case where a boundary vertex $a$ of $C_i$ is the external boundary
vertex and where $a$ does not separate $C_i$ from any part of the
underlying tree. In that case no outside cluster $X_a$ is
associated with $a$.  This completes our implementation of Theorem
\ref{thm:non-localsearch} when there are less than two external
boundary vertices.

\subsection{Two external boundary vertices}
The non-local search described above works fine with less than two
boundary vertices.  However, when we have two external boundary
vertices $x$ and $y$ in the underlying tree $T$, the goal of the
non-local search is to select an edge on $x\cdots y=\pi(T)$. 
In the
above selection process, this means that the currently selected
cluster $C_i$ should always have an edge $e$ from $x\cdots
y$. Then $e\in \pi(C_i)\subseteq \pi(T)$. Thus it follows
that if a child of $C_i$ is not a path child, then that child cannot
be $\Select$ed. In that case, the only path child is automatically made
the next current cluster $C_{i+1}$. The process stops when $\pi(C_i)$
consists of a single edge, which is then returned.

In the actual implementation, since $C_i$ has an edge in its cluster
path, $C_i$ has two distinct boundary vertices $a$ and $b$ with
disjoint outside root clusters $X_a$ and $X_b$. Each of these outside
root clusters contain one of the two external boundary vertices. Let
$A_i$ and $B_i$ be the children of $C_i$ with $a\in A_i$ and $b\in
B_i$. If $A_i$ is not a path child, we
simply set $X_a=\Join(X_a,A_i)$ and $C_{i+1}=B_i$. Similarly, if $B_i$ 
is not a path child,  we set $X_b=\Join(X_b,B_i)$ and $C_{i+1}=A_i$. It
is only if both $A_i$ and $B_i$ are path children that we call
the application-defined $\Select$ on $\Join(A_i^*,B_i^*)$ where
$A_i^*=\Join(X_a,A_i)$ and $B_i^*=\Join(X_b,B_i)$.

We note that with two external boundary vertices $x$ and $y$, it is
necessary that we restrict $\Select$ to pick edges from $x\cdots y$ as
above. Otherwise, above we could end up with $A_i$ and $B_i$
intersecting in a vertex $c$ outside $x\cdots y$.  Since $A_i^*$ and
$B_i^*$ intersect in $c$ and partition the underlying tree, one of
them would contain both $x$ and $y$, hence have three boundary
vertices $x$, $y$, and $c$.

This completes our implementation of Theorem \ref{thm:non-localsearch}.

\section{Methodological remarks}\label{lem:method}
Our results on diameters, centers, and medians could also have been
achieved based on either Sleator and Tarjan's dynamic
trees~\cite{ST83}, or Frederickson's topology
trees~\cite{Frederickson85,Frederickson97}. However, we claim that the
derivation from these more classical data structures would have been
more technical.

\subsection{Frederickson's topology trees} Top trees are very similar to
Frederickson's topology trees~\cite{Frederickson85,Frederickson97},
from which they are derived.  The essential difference is that the
clusters of topology trees are not connected via vertices, but via
edges. Since Frederickson's boundary consists of edges, he cannot
limit the boundaries for unlimited degree trees.  Thus, in
applications for unbounded degrees one has to code these with ternary
trees, inserting some extra edges and vertices that typically require
special handling.  Even if we assume we are dealing with ternary
trees, topology trees still have clusters with up to three boundary
edges instead of just two boundary vertices. Also topology $\Join$
combines two clusters \emph{plus} the edge between them whereas a top
$\Join$ just unites two neighboring clusters.  Neither of these issues
lead to fundamental difficulties, but, in our experience, they lead to
significantly more cases.

We note that Frederickson \cite{Frederickson972} has already shown how
Sleator and Tarjan's \cite{ST83} axiomatic interface to dynamic trees
can be implemented with topology trees. Our corresponding
implementation with top trees from \S~\ref{sec:toptrees} is
inspired by that of Frederickson.

\subsection{Sleator and Tarjan's dynamic trees} Sleator and Tarjan 
provide an axiomatic interface for their dynamic trees~\cite{ST83}
where an application can choose a root with a so-called $\Evert$
operation, and then, for any specific vertex, add weights to all edges
on the path to the root, or ask for the minimum of all weights on this
path. This is basically the interface we implemented with top trees at
the end of \S~\ref{sec:toptrees}, assuming that we expose both
the desired root and the specified vertex. 

Before discussing limitations to the above interface, we first
illustrate its generality by viewing the min-query as representing an
arbitrary associative operator $\oplus$. For example, suppose 
as in~\cite{ST83} that we want to implement parent pointers to the
current root. We then let the weight of an edge be its pair of
end-points and define $a\oplus b=a$.  Then the ``min''-query returns
the end-points of the first edge on the path to the root, from which
we immediately get a parent pointer. Similarly, adding $x$ to all
weights on a path could be done with any associative operator
$\otimes$ that distribute over $\oplus$, that is, $x\otimes(y\oplus
z)=(x\otimes y)\oplus (x\otimes z)$.  Instead of having
$(\oplus,\otimes)=(\min,+)$, we could have e.g.\ 
$(\oplus,\otimes)=(+,\times)$.

Despite these generalizations, the axiomatic interface is
still centered around paths, and it has been found too limited for
many applications of dynamic trees. Instead authors have had to work
directly with Sleator and Tarjan's underlying
representation~\cite{WT92,BT90,Poutre91,Poutre92,Poutre94,GI91,BT89,KTBC91,GGT91,CT97,CT95,CT91,Peckham89}.
In particular, this is the case for the previous solutions to the
dynamic center~\cite{CN96} and median problems~\cite{APP96}, and we
believe part of the reason for their worse bounds and more complex
solutions is difficulties in working directly with Sleator and
Tarjan's underlying representation.

Of course, one may try to increase the applicability of the axiomatic
interface by augmenting it with further operations. For
example,~\cite{Radzik} shows how to find a minimum weight vertex in a
subtree.  However, dealing with non-local properties is not so
immediate, and we find it unlikely that we will ever converge to a set
of operations so big that we can forget about the underlying
representation.

For contrast, with top or topology trees it is easy to
deal directly with the representation. For
example, to compute the minimum vertex of a given subtree as
in~\cite{Radzik}; since we can insert and delete edges, this is
equivalent to maintaining the minimum vertex of each tree in a dynamic
forest. With top trees this is done by maintaining, for each cluster, the
minimum weight over its non-boundary vertices. Since each vertex is
only non-boundary in $O(\log n)$ clusters, weight changes of vertices
are trivially supported. If we do not expose any external boundary
vertices, the root cluster will store the desired minimum.

\subsection{Henzinger and King's ET-trees} For completeness, we also
mention Henzinger and King's ET-trees~\cite{HK95}. This is a standard
binary trees over the Euler tour of a tree. This technique is much
simpler to implement than those mentioned above, and it can be used
whenever we are interested in maintaining a minimum over the edges or
vertices of a tree, where the minimum may be interpreted as any
associative and commutative operation. Thus, the above mentioned
result from~\cite{Radzik} on maintaining the minimum weight vertex of
a tree is immediate, and in fact, this was pointed out
before~\cite{Radzik} in~\cite{Tarjan95}.  However, the ET-trees cannot
be used to maintain any of the path information discussed so far.
Also, they cannot be used to maintain medians and centers.

\section{Generalizations of top trees}\label{gen}
In the following, to avoid confusion with leaves in the underlying
trees, we refer to the leaves of a top tree as {\em base clusters}.
At present the base clusters are just the edges of the underlying
tree, but it is sometimes important to deal with fewer but larger base
clusters. For example, this is needed in classical topology
tree applications such as maintaining the minimum spanning tree of a
fully-dynamic graph~\cite{Frederickson85}. Also, it is needed for a
recent application of top trees maintaining minimum cuts~\cite{Tho01}.
For these applications, we allow the user to distribute \emph{labels}
on the vertices of the underlying tree.  These labels represent
application-specific information associated with the vertices. For
example, if we are maintaining a minimum spanning tree, the labels
represent incident ends of non-tree edges. 

We note that Frederickson's topology trees \cite{Frederickson85,Frederickson97}
do not support labels. His underlying trees have to be ternary so
each application has to decide how to code high degree vertices and other
information in ternary trees.

Thus our top trees are now dealing with a labeled tree $T$. Each label
is attached to a unique vertex, but the same vertex may have
many labels attached. In many regards, the labels can be thought
of as edges with a single end-point. 

In a subtree $U$ of a labeled tree $T$, each vertex may have attached
any subset of its labels in $T$. We extend the notion of boundary
vertices to include vertices in $U$ that have fewer labels attached in
$U$ than in $T$. That is, $\boundary U$ is now the set of vertices in
$U$ that are either external boundary vertices of $T$ or vertices with an incident edge or attached label that is included in
$T$ but not in $U$.

A cluster $U$ of $T$ is a subtree with at most two boundary vertices
containing at least an edge or a label. Thus, we now accept
a single vertex as a cluster if it has an associated
label in the cluster. Two clusters are neighbors if their
intersection is a single vertex. They cannot have
any labels or edges in common. It follows that
the base clusters of a top tree form a partitioning of the
edges and labels of the underlying tree. Similarly, it follows that
labels, like edges, appear in exactly one cluster on each level in a top tree.

One conceptual advantage to labels is that any cluster can be
be reduced to an edge or a label. More precisely, we get a new
labeled tree if we replace a point cluster with a label at its boundary
vertex, or if we replace a path cluster with an edge between its
boundary vertices.

A simple application of labels would be to attach a label $[v]$ to a
vertex $v$. On each level of a top tree, the label $[v]$ will only
appear once whereas the vertex $v$ can participate in arbitrarily many
clusters.  This way, $[v]$ can be used as a distinguished
representative for $v$ in a top tree.

In addition to the original $\Link$, $\Cut$, and $\Expose$ operations,
we have the two new operations:
\begin{procedures}
\item[$\Attach(v,a)$] attaches a label $a$ to the vertex $v$.
\item[$\Detach(a)$] detaches the label $a$ from whatever
vertex it was attached to.
\end{procedures}
To get the full power of the generalized top trees, we allow top nodes
$C$ with a single child $D$, created by $C:=\Join(D)$.  Then $C$ and
$D$ represent exactly the same cluster. We can then get {\em leveled
top trees\/} where all base clusters are on level $0$, and where the
parent of a level $i$ top node is on level $i+1$.  We define the {\em
size of a cluster or labeled tree\/} to be the total number of its
edges and labels. We now have the following generalization of Theorem
\ref{thm:top-maintain}:
\begin{theo}\label{thm:weighted-top-maintain}
  Consider a fully-dynamic forest and let $Q$ be a positive integer
  parameter. For the trees in the forest, we can maintain a leveled
  top trees whose base clusters are of size at most $Q$ and such that
  if a tree has size $s$, it has height $h=O(\log s)$ and $\lceil
  O(s/(Q(1+\varepsilon)^i))\rceil$ clusters on level $i\leq h$.  Here
  $\varepsilon$ is a positive constant.  Each $\Link$, $\Cut$,
  $\Attach$, $\Detach$, or $\Expose$ operation is supported with
  $O(1)$ $\Create$s and $\Destroy$s, and $O(1)$ $\Join$s and $\Split$s
  on each positive level.  If the involved trees have total size $s$,
  this involves $O(\log s)$ top tree modifications, all of which are
  identified in $O(Q+\log s)$ time. For a composite sequence of $k$
  updates, each of the above bounds are multiplied by $k$.
  As a variant, if we have parameter $S$ bounding the size of each 
  underlying tree, then we can choose to let all top roots be on the
  same level $H=O(\log S)$.
\end{theo}
We note that Theorem~\ref{thm:weighted-top-maintain} implies
Theorem~\ref{thm:top-maintain}. More precisely, to get
Theorem~\ref{thm:top-maintain} from Theorem~\ref{thm:weighted-top-maintain}, 
we set $Q=1$, use no labels, and skip all top nodes that are single children.

To appreciate Theorem~\ref{thm:weighted-top-maintain}, we briefly
sketch Frederickson's algorithm for maintaining a minimum spanning
tree of a fully-dynamic graph~\cite{Frederickson85}, but using top
trees instead of topology trees. 
\begin{theo}[Frederickson] We can maintain a minimum spanning tree of a 
  fully dynamic
  connected graph in $O(\sqrt{m})$ time per edge insertion or 
  deletion\footnote{We note that for denser graphs, 
  Eppstein et al.~\cite{EGIN97} have 
  improved the $O(\sqrt{m})$ bound to $O(\sqrt{n})$ using their general 
sparsification technique.}.
\end{theo}
\begin{pf}
  If an edge $(v,w)$ is inserted in the graph, it should be added to
  the minimum spanning tree $T$ if it is lighter than the maximum
  weight on the path from $v$ to $w$ in $T$. From
  Theorem~\ref{cor:path-query}, we already know how to support such
  path queries in $O(\log n)$ time.
  
  Our challenge is to deal with the deletion of a tree edge. Our task
  is to find a lightest replacement edge reconnecting the tree, and we
  will show how to do this in $O(\sqrt{m})$ time. 
  
  We will employ leveled top trees $\Top$ from
  Theorem~\ref{thm:weighted-top-maintain} where the labels attached to
  a vertex are ends of incident non-tree edges. More precisely, for
  each non-tree edge $(v,w)$, we have a label $[v,w]$ attached to $v$
  and a symmetric label $[w,v]$ attached $w$.  These two labels are
  always attached or detached as a composite update (c.f. \S
  \ref{top-dicipline}) so that we never have one but not the other
  present in our top trees. The total size of our labeled forest
  is then the number $m$ of edges in the graph.

  We will use the variant of top trees in the end of
  Theorem~\ref{thm:weighted-top-maintain} with $S$ an upper bound on
  the total size $m$. Using standard back-ground rebuilding, we can
  ensure $S=\Theta(m)$. More precisely, we can divide updates into
  epochs that first initiate new top trees $\Top'$ with this $S'=2m$
  instead of the current $S$. During the next $S/4$ updates we copy
  the current data from $\Top$ to $\Top'$, and switch to $\Top'$ when done.

  Now that $S$ is fixed for the current top tree $\Top$, we set
  $Q=\sqrt{S}=\Theta(\sqrt{m})$. Since we have at most two trees
  at any time, the number of clusters on level $i\leq
  H=O(\log S)=O(\log m)$ is $\lceil
  O(S/(Q(1+\varepsilon)^i))\rceil=\lceil O(\sqrt
  m/(1+\varepsilon)^i)\rceil$.

  For each pair $(C,D)$ of clusters on the same level,
  we will store the lightest non-tree edge $lightest(C,D)$ between
  them. Here $(v,w)$ goes between $C$ and $D$ if $[v,w]$ is a label in
  $C$ and $[w,v]$ is a label in $D$, or vice versa. Assuming that
  the clusters are enumerated with numbers up to $O(\sqrt S)$, we can 
  implement $lightest$ as a simple two dimensional array over
  all cluster pairs. We can just ignoring entries with cluster
  pairs on different levels. Also, since $lightest$ is symmetric,
  we identify $lightest(C,D)$ with $lightest(D,C)$.
  
  Assuming that the array $leightest$ is properly maintained, 
  if a tree edge $(v,w)$ is deleted, we $\Cut$ it, and
  then the desired minimum replacement edge is the minimum edge
  between the root clusters. More precisely, we perform the following
  sequence of operations: 
\begin{eqnarray*} &&\Cut((v,w));\
  C:=top\_root(v);\ D:=top\_root(w);\ (x,y):=lightest(C,D);\\
  &&\langle \Detach([x,y]);\ \Detach([y,x]);\rangle \; \Link((x,y));
\end{eqnarray*} 
  We now have to show how to maintain $leightest$. 
  Suppose a base cluster $B$ is
  $\Create$d.  Since it has only $\sqrt{m}$ incident non-tree edges,
  each going to a base cluster on the same level, we can easily find
  $lightest(B,D)$ for all the $O(\sqrt{m})$ base clusters $D\in\Top$
  in $O(\sqrt{m})$ time.
  
  Now suppose a level $i>0$ cluster $C$ is $\join$ed.  For each of the
  $\lceil O(\sqrt{m}/(1+\eps)^i)\rceil$ other level $i$ clusters
  $D\in\Top$, we set $lightest(C,D)$ to be the lightest of
  $lightest(A,B)$ where $A$ is a child of $C$ and $B$ is a child of
  $D$. Thus we compute $lightest(C,D)$ in constant time.

  Finally, we note that $\Split$ and $\Destroy$ require no action.  It
  follows from Theorem~\ref{thm:weighted-top-maintain} that each
  $\Link$, $\Cut$, $\Expose$, $\Attach$, or $\Detach$ operation is supported in 
  \[O(\sum_{i=0}^{O(\log n)}\lceil\sqrt{m}/(1+\eps)^i\rceil)=O(\sqrt m)\] 
  time, which is then also the time bound for
  finding a replacement edge.
\end{pf}

A much more involved application using the generalized top trees from
Theorem~\ref{thm:weighted-top-maintain} is the fully-dynamic algorithm
for maintaining minimum cuts~\cite{Tho01}. We note that~\cite{Tho01}
assumes Theorem~\ref{thm:weighted-top-maintain} which is proved below in
this paper by reduction to Frederickson's topology trees \cite{Frederickson97}.

\section{Implementing top trees}\label{sec:implement}
We will now first implement the top trees of
Theorem~\ref{thm:weighted-top-maintain} via Frederickson's topology
trees~\cite{Frederickson97}, and thereby establish
Theorem~\ref{thm:weighted-top-maintain} and
Theorem~\ref{thm:top-maintain}.  Next, we implement the topology trees
with Sleator and Tarjan's st-trees~\cite{ST83}. The connection is
interesting because topology trees and st-trees so far have been
implemented with very different techniques. A nice consequence is that
the simple amortized implementation of st-trees implies a simple
amortized implementation of topology trees, and of top trees.
Previously, no simple amortized implementation of topology trees was
known. We note that for a practical implementation, one should not
follow all our reductions rigorously, but rather go for a more direct
implementation. We hope to address these practical issues in future work.

\subsection{Implementing $\Expose$}
As a very first step in our reduction, we note that if we first have
an implementation of top trees without $\Expose$, then later,
we can easily add $\Expose$. The simple point is that
in a top tree of
height $h$, each vertex is included in at most $h$ clusters. To
$\Expose$ $a$ and $b$, we simply $\Split$ all the clusters having them
as non-boundary vertices.  We now have a set of $O(h)$ root clusters
to be $\join$ed into one cluster. Clearly, this can require at most
$O(h)$ $\Join$s, so we do not need to worry about the height. First,
as long as there is a point cluster, we $\Join$ it with an arbitrary
neighbor. If $a=b$, this process ends with a single point cluster, as
desired. Otherwise, we end with a string of path clusters
$C_1,...,C_k$ with boundaries $\Set{c_0,c_1}$, $\Set{c_1,c_2}$, ...,
$\Set{c_{k-1},c_k}$ where $c_0=a$ and $c_k=b$.  We can then repeatedly
$\Join$ neighbors in this string until a single path cluster with
boundary $\Set{a,b}$ remains. Before supporting any new $\Link$ or
$\Cut$, we simply revert all the above $\Join$s and $\Split$s,
restoring the previous un-exposed top tree.

Thus, in the remaining implementation, we may consider
$\Expose$ done, and focus on maintaining top trees of height $O(\log n)$ 
under $\Link$ and $\Cut$ as in Theorem~\ref{thm:weighted-top-maintain}
but without $\Expose$.

\subsection{Top trees via topology trees}
Theorem~\ref{thm:weighted-top-maintain} without $\Expose$ is proved
in~\cite{Frederickson97} in the context of topology trees with their
different definition of clusters.
The topology clusters are subtrees like top clusters,
but in a topology tree, independent clusters  are vertex-disjoint. 
In particular, the topology base clusters are
disjoint. They partition the vertices and are connected via edges. The
topology trees are only defined for ternary trees. A cluster may have
at most 3 edges leaving it, called boundary edges, and if it has
three edges leaving it, it may only consist of a single vertex. The
topology tree is binary like a top tree.  A parent cluster is the
union of the two child clusters plus the edge connecting them.

Now, implementing top trees with topology trees is easy. We ternarize
each vertex as follows: while there is a vertex $v$ with degree $>3$,
we turn $v$ into a path with the incident edges branching off.  More
precisely, if $v$ is incident to $w_0,...w_d$, $d\geq 3$, we may
replace $v$ by a path $v_1,...,v_{d-1}$ with incident edges
$(v_1,w_0)$, $(v_i,w_i)$, $i=1,...,d-1$, and $(v_{d-1},w_d)$. The edge
$(v_i,w_j)$ remembers that it originated from $(v,w_j)$. In
Frederickson's topology trees the base clusters are all disjoint. To
represent labels associated with a vertex $v$, we just add them to the
above path representing $v$ as extra vertices.

To transform a topology tree into a top tree, we essentially just take
each topology cluster $C$ and transform it into the top cluster $C'$
induced by the vertices, edges, and labels contained in $C$.  We note that $C'$
has at most two boundary vertices. Clearly this is the case if $C$ has
at most two boundary edges, but if $C$ has three boundary edges, $C$
consists of a single vertex, which is hence the only boundary vertex.
As an exception, if a topology cluster has no labels or edges from
the original tree, there is no corresponding top cluster is considered
empty and has no representative in the top tree.

The base top clusters are those derived from the base topology
clusters, plus a base cluster for each edge not in a derived base
cluster.  Now, a topology $\Join$ in converts into two top $\Join$s,
where first one of the topology children $\Join$ with the edge between
them.  Next the resulting top cluster $\Join$s with the other topology
child. Here a $\Join$ with an empty top cluster is just skipped. Since
a topology $\Join$ may requires two top $\Join$s, each level in a
topology tree translates into two levels in a top tree. Given the
proofs for topology trees in~\cite[pp. 486--497]{Frederickson97}, we
conclude that Theorem~\ref{thm:top-maintain}
and~\ref{thm:weighted-top-maintain} hold true. The achievement with
top trees is a simpler interface for high-degree trees where the
ternarization is not done by each application but by the
implementation via the above reduction. Also, the $\Join$ has slightly
fewer cases and is slightly simpler because we do not have to
incorporate an edge between the clusters.

\subsection{Topology trees via st-trees}
We will now demonstrate how Sleator and Tarjan's st-trees~\cite{ST83}
can be used to implement topology trees whose base clusters are the
vertices. Together with the previous reduction from top trees to 
topology trees, this provides us with a very different implementation of 
Theorem~\ref{thm:top-maintain}.
Here by st-trees, we do not refer to the nice 
path-oriented axiomatic interface from~\cite{ST83}, but to the 
underlying implementation.

First, we note that the st-trees are presented for rooted trees, but
on the other hand, they have an $\Evert(v)$ operation, making $v$ the
root of its tree. Hence, to perform an arbitrary $\Link(u,v)$, we can
first $\Evert(u)$, making it root of its tree, and then $\Link(u,v)$,
making $(u,v)$ a parent pointer.

Since our starting point is an unrooted ternary tree, a rooted version of 
it is a binary tree. An exception is the root, which in principle could be
have three children. However, this is easily avoided. First of all,
we could pick the root as a a leaf in the unrooted
tree with degree one. Also, consider the situation above where we want to 
$\Link(u,v)$ and first make $u$ the root with $\Evert(u)$. Since the result is
ternary, $u$ had degree at most two before $\Link(u,v)$, so $u$ does not
get three children. The $\Link(u,v)$ operation is just adding a parent
pointer to $u$.

Sleator and Tarjan define a set of disjoint solid paths
down from a vertex in $T$ to a leaf providing a partitioning of the
vertices. They then form an st-tree $T'$ as follows. They take each
solid path $P=(v_1,...,v_p)$ with $v_1$ closest to the root and $v_0$
the parent of $v_1$, and remove all parent pointers of the vertices in
the path.  Then they make a binary tree $P'$ with $v_1,...,v_p$ as
leaves appearing in this order, and make $v_0$ the parent of the
root. If $v_1$ was the root of the whole tree, the root of $P'$
becomes the root of $T'$, which in~\cite{ST83} ends up with
logarithmic height.

Now each vertex $v$ in $T'$ represents the cluster $C(v)$ induced by
the vertices from $T$ descending from it in $T'$.  To see that these
are clusters we just note that if $v\in P'$ above, the descendants of
$v$ from $P$ form a segment $S$ of $P$. The only edges incident to
$C(v)$ are then the parent pointer from the first vertex in $S$ and
the children pointer from the last vertex in $S$ to its child in $P$,
if any.

We can now construct the topology tree as follows. The base clusters
are the vertices of $T$. The rest of the top tree is constructed by
following $T'$ bottom-up. When we meet a vertex $v$ from $T$, it has
only one child $w$ in $T'$, which was its non-solid child in $T$. Then
$C(v)=\Join(\Set{v},C(w))$. When we meet a vertex $v'$ not from $T$,
it has two children $u$ and $w$ in $T'$, and then
$C(v')=\Join(C(u),C(w))$.

Thus we have established a mapping from the st-tree $T'$ to a topology
tree $\Top$ whose base clusters are the vertices. Since the st-tree has height $O(\log n)$ so
does the topology tree. Also, the main technical result
from~\cite{ST83} is that each $\Link$, $\Cut$, and $\Evert$, only
affects $O(\log n)$ vertices in the st-trees, including their parents,
and hence this gets translated into $O(\log n)$ $\Split$s and
$\Join$s.  Thus, we can derive Frederickson's topology trees
~\cite{Frederickson97}, and hence top trees, from Sleator and Tarjan's
st-trees~\cite{ST83}. In particular this implies that the simple
amortized version of st-trees ~\cite{ST85} provides a simple amortized
version of top trees. When using the amortized version of top
trees, there is no guarantee of the height of the top tree. However,
if we precede each query with an $\Expose$ we will meet the amortized
bounds.

The advantage of top trees and topology trees over st-trees is a
nice, easy to apply, interpretation of the system of solid paths
replaced by binary trees in st-trees. This point is illustrated with
our top tree solutions to the diameter, center, and median problems
for dynamic trees, improving over previous solutions based on st-trees
~\cite{APP96,CN96}.

\section{Concluding remarks}\label{sec:conclussion}
We have introduced top trees as a {\em design\/} of an interface providing 
users with easier access to the power of previous techniques for maintaining
information in a fully-dynamic forest. Conceptually, top trees are
very similar to Frederickson's topology trees \cite{Frederickson97}, 
the subtle difference
being that top clusters are $\Join$ed by vertices whereas topology trees
are $\Join$ed via edges. This small difference has the immediate advantage
that top trees work directly for trees of unbounded degrees, which with
topology trees would first have to be coded as ternary trees. It
also makes $\Join$s of two clusters a bit simpler in that they do not 
involve an intermediate edge. 

Using top trees, we dealt with a variety of different applications including
non-local search problems like maintaining the center or median of
trees in a dynamic forest. For these two problems, we provided quadratic
improvements over previous bounds.
We also showed how top trees, in theory, could be implemented both
with Frederickson's topology trees \cite{Frederickson97}, and with 
Sleator and Tarjan's st-trees~\cite{ST83}. 

A main practical challenge is now to make a good library implementation
of top trees for use in different applications.  We could have
different implementations, e.g., a worst-case implementation based on
the ideas in topology trees \cite{Frederickson97}, and a faster
amortized implementation based on st-trees~\cite{ST83}.  For speed,
the implementations should be tuned directly for top trees and not
just use our general reductions. Ideally, applications and
implementations should only communicate with each other via the top
tree interface, so that one can replace one implementation with
another in a plug-and-play manner without a change to the
applications. It is not trivial to make such generic interfaces
efficient, but C++ solutions have been reported in by Austern et
al.~\cite{ASTW03} for the simpler case of balanced binary search
trees. We do hope to address such practical library implementations of
top trees in future work.

\section*{Acknowledgment} We would like to thank Renato Werneck
for many very helpful comments to an earlier version of this paper.


\end{document}